\documentclass[aps, pre, twocolumn, superscriptaddress, amsmath,amssymb,floatfix]{revtex4}

\newif\ifpdf\ifx\pdfoutput\undefined\pdffalse\else\pdfoutput=1\pdftrue\fi

\newcommand{\figurewidth}{0.46\textwidth}

\usepackage{graphicx}
\usepackage{bm}
\usepackage{epsfig}

\newcommand{\be}{\begin{equation}}
\newcommand{\ee}{\end{equation}}

\newcommand{\old}{u}
\newcommand{\new}{v}

\begin{document}

\title{\bf Grand canonical simulation of phase behaviour in highly size-asymmetrical binary fluids}

\author{Douglas J. Ashton}
\author{Nigel B. Wilding}
\affiliation{Department of Physics, University of Bath, Bath BA2 7AY, United Kingdom.}

\begin{abstract}

We describe a Monte Carlo scheme for the grand canonical simulation
study of fluid phase equilibria in highly size-asymmetrical binary
mixtures. The method utilizes an expanded ensemble in which the
insertion and deletion of large particles is accomplished {\em gradually} by
traversing a series of states in which a large particle interacts only
partially with the environment of small particles. Free energy barriers
arising from interfacial coexistence states are surmounted with the aid
of multicanonical preweighting, the associated weights being determined
from the transition matrix. As an illustration, we present results for
the liquid-vapour coexistence properties of a Lennard-Jones binary mixture
having a $10:1$ size ratio. 

\end{abstract}

\maketitle

\section{Introduction}

Fluid mixtures comprising two or more particle species of disparate
sizes are common in soft condensed matter \cite{BELLONI00}. A prime
example is a colloidal dispersion to which much smaller particles have
been added such as non-absorbing polymers
\cite{russel1989,POON02,Zaccarelli2009} or charged nanoparticles
\cite{LIU2005_1}. Interest in such systems stems from the fact that by
judicious choice of the small component, one can potentially control the
equilibrium and dynamical properties of the large component, giving rise
to a rich assortment of novel phenomena and material properties
\cite{Belloni2000,russel1989}. Given, however, the wide variety of small
particles that one might conceivably choose to add, the experimental
task of characterizing the range of possible behaviour is considerable.
With this in mind there has been much interest in deploying statistical
mechanics and computer simulation to {\em predict} the properties of
such mixtures. 

In this paper we shall focus on the problem of obtaining the equilibrium
phase behaviour of models of highly size-asymmetrical mixtures. Direct
analytical assaults on such systems are generally complicated by the
disparity in particle length scales \cite{Ayadim2006}. To make progress, a widely
practiced simplifying strategy is to try to map the true two-component
mixture onto a single component system comprising solely the colloid
particles. These are assumed to interact via an {\em effective}
potential which is supposed to represent the net effect of the bare
colloid-colloid interactions plus the additional interactions mediated
by the small particles. Arguably the most successful example of such an
approach pertains to particles that interact as hard spheres  -- a
situation which can be realized experimentally to a good approximation
in colloid-polymer mixtures \cite{Pusey}. Here the effective
interaction is the celebrated ``depletion'' potential describing the
interaction between two hard sphere colloids immersed in a ``sea'' of
small hard spheres \cite{Asakura1954}. In seminal work, Bob Evans and
coworkers have contributed much insight into this situation by tracing
out the degrees of freedom associated with the small particles in order
to produce an explicit expression for the depletion potential
parameterized by the particle size ratio and the volume fraction of
small particles. This not only provides valuable information on the
nature of the colloidal interactions, but also serves as a basis for
theoretical and simulation investigations of the phase behaviour of the
effective one component system \cite{GOETZELMANN98,Dijkstra1998,Dijkstra1999}.

Whilst impressive progress has been made in obtaining accurate effective
one-component potentials, at present they are largely limited to
underlying interactions of the hard sphere form \cite{BELLONI00}.
Moreover, because effective potentials are usually derived in the limit
of low density of large particles, there are concerns about their
accuracy at high densities where many body effects are significant.
Ideally then, one should like to be able to tackle the full two
component system and treat arbitrary interactions between the particle
species. Achieving this analytically still seems some way off, making it
tempting to appeal to computer simulation for help. Unfortunately,
simulations of highly size asymmetric mixtures encounter their own
problems: the relevant physics is controlled by the length scale of the
large particles, but attempts to relax these particles are often
frustrated by the presence of the small particles. For instance grand
canonical Monte Carlo simulations -- the method of choice for studies of
fluid phase transitions \cite{Wilding1995} -- suffer an unfeasibly small
acceptance rate for insertions of large particles. Similarly in
Molecular Dynamics an impractically small timestep is mandated by the
need to avoid high energy overlaps between large and small particles.

In this paper we describe a tailored Monte Carlo simulation scheme that
circumvents the principal drawbacks of traditional approaches. The
essential idea is to treat both species grand canonically, but to ease
the sampling bottleneck for insertions (and deletions) of large
particles by performing these -- not in a single Monte Carlo step -- but
{\em gradually}. In practice this is achieved by permitting the system
to traverse (in a stochastic fashion) a prescribed set of states (or
``stages'') that interpolate between the limits of a large particle being fully
present and fully absent from the system. This idea of staged insertion
has been around for some time, principally in the context of chemical
potential measurements for dense fluids and complex molecules using the
Widom formula
\cite{Mon1985,Nezbeda1991,Attard1993,Kaminsky1994,wilding1994a,Frenkelsmit2002}.
It has been recently revisited in the context of optimizing expanded
open ensembles by Escobedo \cite{Escobedo2007}. However, to our
knowledge it has not been used to calculate the full phase behaviour of
a model asymmetric mixture at large ratios of the component sizes.

\section{Method}

In this section we begin by outlining the statistical mechanical basis
to the staged insertion method for a highly size asymmetric binary
mixture. Thereafter we discuss implementation issues, taking as an
example the case of a Lennard-Jones (LJ) mixture.

\subsection{Statistical mechanics} \label{sec:statmech}

Consider a binary mixture comprising $N$ particles, $N_l$ of which are `large'
(l) and $N_s$ of which are `small' (s), all contained in a volume $V$ at
temperature $T$. Particles are identified via an index $1\le i\le
N$, and a species label $\gamma_i=l,s$, and we write the internal energy as

\be
\Phi=\sum_{i=1}^{N}\sum_{j=i+1}^{N}\phi_{\gamma_i,\gamma_j}({\bf q}_i,{\bf q}_j)\;,
\ee
where $\phi_{\gamma_i,\gamma_j}$ is the pair potential for particles $i$
and $j$ of species $\gamma_i$ and $\gamma_j$, located at position vectors
${\bf q}_i,$ and ${\bf q}_j$ respectively. 

\begin{figure}[h]
  \includegraphics*[width=\figurewidth]{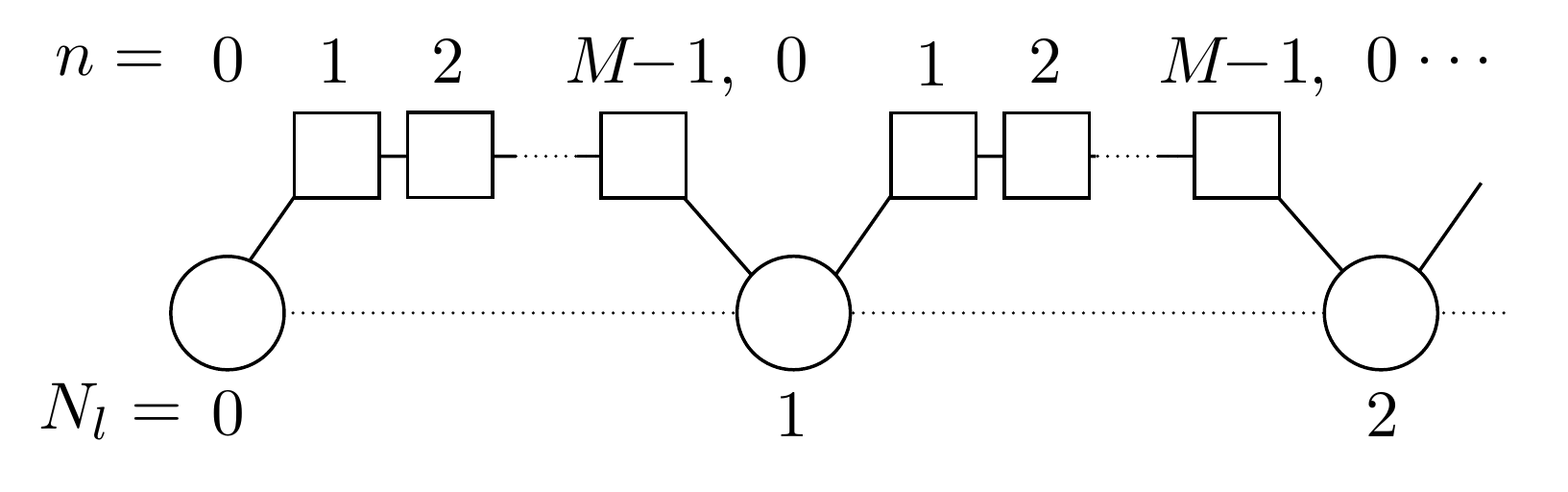}
\caption{Schematic showing how each integer value of
the large particle number $N_l$ is expanded into $M$ 
stages, each of which is indexed by the macrovariable $n$.}
\label{fig:stages}
\end{figure}

Let us now augment this system with an additional `ghost' ($G$) large
particle having position vector ${\bf q}_G$. The ghost particle is
taken to interact normally with other large particles, but differently
with small particles. To deal with this, it is more convenient to associate separate indices
$k$ and $m$ with the $N_l$ large and $N_s$ small particles
respectively, and write the interaction of the ghost particle as

\be \label{eq:phig}
\Phi_G=\sum_{k=1}^{N_l}\phi_{ll}({\bf q}_k,{\bf q}_G)+\sum_{m=1}^{N_s}\tilde\phi_{ls}^{(n)}({\bf
q}_m,{\bf q}_G)~.
\ee
Here $\tilde\phi_{ls}^{(n)}$ describes the interaction between the ghost
large particle and a small particle. This is modified with respect to
the standard large-small interaction by the dependence on a discrete
stochastic macrovariable $n=0\ldots M-1$. The role of $n$ is to index the
stages that specify the degree of coupling between the ghost and the small
particles. Fluctuations in $n$ forwards or backwards across its range
result in the gradual insertion or deletion of a large particle (Fig~\ref{fig:stages}). To be
more specific, we let $n=0$ correspond to $N_l$ large particles, while
$n=M$ corresponds to $N_l+1$. Intermediate values of $n=1\ldots M-1$
represent a system of $N_l$ large particles plus a ghost particle. Thus
transitions $n=1\rightarrow 0$ correspond to the deletion of the ghost particle
from the system, while $n=M-1\rightarrow M$ correspond to it turning into a fully
interacting (ie. standard) large particle.  In this sense the $n=M$ state
for a system of $N_l$ large particles and the $n=0$ state for a system
of $N_l+1$ large particles are equivalent.

The internal energy of the augmented system is $\Phi^\prime(\{{\bf q}\}_l,\{{\bf q}\}_s,q_G,n)=\Phi+\Phi_G$
and the associated `expanded' \cite{lyubartsev1992} canonical ensemble
has the partition function $Z^\prime(N_l,N_s,V,T,n)$, where

\be
\label{eq:canpart}
Z^\prime=\prod_{k=1}^{N_l}\prod_{m=1}^{N_s}\int d{\bf q}_k \int d{\bf q}_m \int d{\bf q}_G\exp[-\beta\Phi^\prime],
\ee
with $\beta=1/k_BT$. In the present work, we shall be concerned with the measured
form of the {\em grand canonical} (GC) ensemble probability distribution of
the fluctuating number of large particles, $p(N_l|\mu_l,\mu_s,V,T)$,
where $\mu_s$ and $\mu_l$ are the chemical potentials of the small and
large species respectively. This is obtainable from measurements of
the joint distribution $p(N_l,n|\mu_l,\mu_s,V,T)$ conducted within the expanded GC
ensemble, which is defined via a weighted sum of the expanded
canonical ensemble partition function $Z^\prime$:

\be
\label{eq:probNln}
p(N_l,n)\simeq \sum_{N_s=0}^\infty Z^\prime\exp\left[\beta(N_l\mu_l+N_s\mu_s)\right]\;.
\ee
Here $\simeq$ means up to an arbitrary normalization constant and (for brevity) we have omitted combinatorical and volume
factors.  $p(N_l|\mu_l,\mu_s,V,T)$ follows from Eq.~\ref{eq:probNln} by picking out those
macrostates from the expanded ensemble having $n=0$, ie. that
correspond to the physical states in which no ghost particles are
present in the system:

\be
p(N_l)= \frac{1}{{\cal Z}}\sum_{N_s=0}^\infty \sum_{n=0}^{M-1} Z^\prime\exp\left[\beta(N_l\mu_l+N_s\mu_s)\right]\delta_{n,0}\,
\ee
where 

\be
{\cal Z}=\sum_{N_l=0}^\infty\sum_{N_s=0}^\infty\sum_{n=0}^{M-1}Z^\prime\exp\left[\beta(N_l\mu_l+N_s\mu_s)\right]\delta_{n,0}
\ee
is the grand partition function.

In the present work we shall seek to obtain $p(N_l,n)$ at state points
$(\mu_l,\mu_s,T)$ for which its form may vary over many decades.
Variations on such a scale preclude direct measurements of $p(N_l,n)$
unless special biasing techniques are deployed to facilitate
sampling of the regions of intrinsically low probability. To this end we
utilize multicanonical preweighting \cite{berg1992}, specifying a sampling distribution

\be
\hat p(N_l,n|w)\simeq p(N_l,n)\exp[w(N_l,n)],
\label{eq:weighted}
\ee
where $w(N_l,n)$ represents a set of weights defined on the discrete
combinations $\{N_l,n\}$. As shall be described in
Sec.~\ref{sec:implement}, these weights are chosen such as to ensure
approximately uniform sampling on the set. The desired form of $p(N_l)$
is regained from the measured form of $\hat p(N_l,n)$ by first using
Eq.~\ref{eq:weighted} to unfold the effects of the weights, then picking
out those macrostates having $n=0$. 

\subsection{Implementation for a binary Lennard-Jones mixture}
\label{sec:implement}

In order to illustrate how the above formalism can be implemented in
practice, we consider the case of a binary mixture of Lennard-Jones
particles. Pairs of particles labelled $i$ and $j$ (having respective
species labels $\gamma_i$ and $\gamma_j$) interact via the potential 

\begin{equation}  \label{eq:lj}
\phi_{ij}(r) = 4 \varepsilon_{\gamma_i\gamma_j} \left[ \left(\frac{\sigma_{\gamma_i\gamma_j}}{r}\right)^{12} - \left(\frac{\sigma_{\gamma_i\gamma_j}}{r}\right)^6\right]\; .
\end{equation}
Here $\varepsilon_{\gamma_i\gamma_j}$ is the well depth of the
interaction, while $\sigma_{\gamma_i\gamma_j}$ sets the range of the
interaction based on the additive mixing rule
$\sigma_{\gamma_i\gamma_j}=(\sigma_{\gamma_i}+\sigma_{\gamma_j})/2$, 
where  $\sigma_{\gamma_i}$ and $\sigma_{\gamma_j}$ are the  particle
diameters. Interactions are truncated at
$r_c=2.5\sigma_{\gamma_i\gamma_j}$ and we take $\sigma_l$ as our unit
length scale.

We shall be concerned with state points in which the small particles
occupy a relatively small fraction of the overall volume and act
as a quasi-homogeneous background to the large ones. Under
these circumstances, configurations of small particles can readily be
sampled using a standard GC algorithm at constant chemical potential,
$\mu_s$. As is customary (in order to make contact with experimental
scenarios), we choose $\mu_s$ to yield a prescribed volume fraction,
$\eta^r_s$, of small particles in the {\em reservoir} \cite{footnote1}. Since we seek a
quasi-uniform density of small particles, we set
$\varepsilon_{ss}=\varepsilon_{ls} = \varepsilon_{ll}/10$, which ensures
that the small particle reservoir fluid lies well above its own
(liquid-vapour) critical temperature. In the results of Sec.~\ref{sec:results}
we refer to a dimensionless temperature which is defined as
$T^\star=1/(\beta\varepsilon_{ll})$.

For highly size-asymmetric mixtures, a large number of small particles
are typically found within the cutoff radius $2.5 \sigma_{ls}$ of each
large particle. In order to locate efficiently these particles, we
partition our cubic simulation box of volume $V=L^3$ into cubic
cells of linear extent $2.5 \sigma_{ls}$, and maintain a list of cell
occupancies. Similar cells structures were employed to identify small-small
and large-large interactions \cite{footnote2}.

As described in Sec.~\ref{sec:statmech}, a large particle is inserted or
deleted in stages by modifying its interaction with the small particles.
Accordingly one must specify in advance the form of the ghost particle
interaction for each stage $n$. Obvious possible strategies include
varying the well depth of the interaction, or the range. However, we
have found that neither of these approaches operates very effectively in
practice because of the rapid increase of the potential for 
distances less than that of the potential minimum. Specifically, particles
whose separation is such that the interaction energy is small at one
value of $n$ can incur a very high energy penalty at a neighbouring
stage. This impacts adversely on the acceptance rate, a difficulty which
can only be mitigated by employing a large total number of stages $M$. 

\begin{figure}[h]
\includegraphics*[width=\figurewidth]{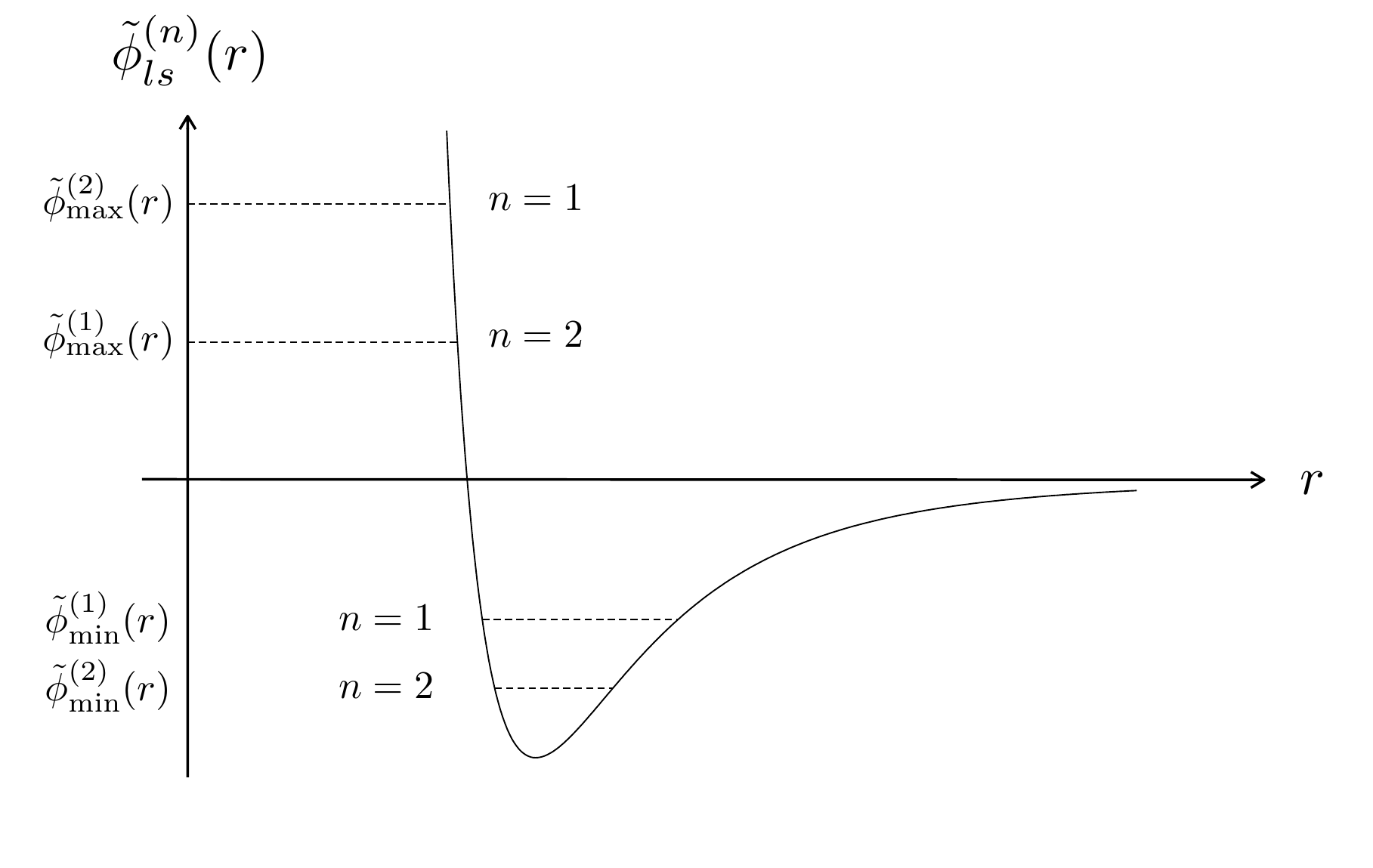}
\caption{Schematic form of the interaction between the ghost particle
and a small particle, $\tilde \phi_{ls}^{(n)}(r)$
(Eq.~\ref{eq:ghostpot}) for two values of $n$, compared to the full
LJ interaction potential between large and small particles. }
\label{fig:intermediates}
\end{figure}

A superior strategy circumvents this problem by imposing a minimum on
the attractive part of the interaction potential and a maximum on the
repulsive part:

\be
\tilde \phi_{ls}^{(n)}(r) = \left\{ 
\begin{array}{l l}
\min(\phi_{ls}(r), \tilde\phi^{(n)}_{\mathrm{max}}) & \quad r<\sigma_{ls}\\
\max(\phi_{ls}(r), \tilde\phi^{(n)}_{\mathrm{min}})  & \quad r\ge\sigma_{ls}\\
\end{array} \right. \;.
\label{eq:ghostpot}
\ee
Each stage, $n$, is thus specified by a pair of parameters,
$\tilde\phi_{\mathrm{min}}^{(n)}$ and
$\tilde\phi_{\mathrm{max}}^{(n)}$. The form of $\tilde
\phi_{ls}^{(n)}(r)$ for two such stages is compared schematically with the
full potential $\phi_{ls}(r)$ in fig.~\ref{fig:intermediates}.

Once the set of stages has been defined, a Monte Carlo scheme for
sampling them can be implemented. Given a system of $N_l$ large
particles and a ghost particle at stage $n$, a proposal is made to
perform a transition to an adjacent stage, $n\rightarrow n^\prime$. This
proposal is accepted or rejected according to a simple Metropolis
criterion

\be
p_{\rm acc} = \min\bigl(1, \exp\bigl[-\beta ( \Phi_G^{(n^\prime)} - \Phi_G^{(n)})
+\Delta w\bigr]\bigr)\;,
\label{eq:pacc}
\ee
where $\Phi_G$ is given by Eq.~(\ref{eq:phig}) and $\Delta
w=w(N_l,n)-w(N_l,n^\prime)$ is the difference in multicanonical weights
in the old and new states, the specification of which is discussed
below. Note that special measures pertain to transitions that bring
the ghost particle to the end of the range of $n$. Specifically, for a
transition $n=1\rightarrow 0$, the ghost particle is completely removed
from the system; the reverse move entails a new ghost being added at a
randomly chosen location. On the other hand, when a ghost particle
undergoes a transition $n=M-1\rightarrow M$, it becomes fully coupled to
the rest of the system, $\tilde \phi_{ls}(r) = \phi_{ls}(r)$, and
$N_l\rightarrow N_l+1$; the corresponding reverse move entails 
nominating a randomly chosen large particle to become a ghost and
setting $N_l\rightarrow N_l-1$. In such circumstances the difference in
weights appearing in Eq.~\ref{eq:pacc} is
$\Delta w=w(N_l,n)-w(N_l^\prime,n^\prime)$.

In standard GC simulation, updates that insert or remove a particle
usually incorporate a factor of $e^{\beta \mu_l} V/(N_l+1)$ (insertion)
or $e^{-\beta \mu_l} N_l/V$ (deletion) in $p_{\rm acc}$ to yield the
correct GC equilibrium distribution. When operating in the expanded GC
ensemble it is convenient (in the interests of obtaining a smooth weight
function in the expanded space of $N_l$ and $n$) to set the chemical
potential $\mu_l=0$ and to ignore the volume and particle number factors
for the time being. The neglected factors, as well as the unfolding of
the multicanonical weights (cf. Eq.~\ref{eq:weighted}) are easily
accounted for when extracting the final GC distribution from the
measured form of $\hat p(N_l,n)$:

\begin{eqnarray}
\log p(N_l | \mu_l) & \simeq & \log \hat p(N_l,n=0|\mu_l=0) + \beta \mu_lN_l  \\
& & - w(N_l,n=0) + N_l\log V - \log (N_l!).\nonumber
\end{eqnarray}

We turn now to the matter of the choices for the number of stages $M$
and the associated values of the stage parameters
$\tilde\phi_{\mathrm{min}}^{(n)}$ and $\tilde\phi_{\mathrm{max}}^{(n)}$.
This is governed by three main desiderata : 
\begin{enumerate}
\item[(i)] The rates for transitions between neighbouring stages should be roughly equal (in both
directions) in order to avoid bottlenecks in the sampling. 
\item[(ii)] $M$ should be sufficiently large to ensure a reasonably high transition rate.
\item[(iii)] The number of stages $M$ should not be so large that the correlation
time of the resulting random walk in $\{N_l,n\}$ is excessive (bearing
in mind that the time to cover a given number of steps grows like the
square of the number of steps). 
\end{enumerate}

With regard to (i), as we have chosen to implement it, staging solely
influences the strength of interaction between the ghost large particle
and the small particles. Hence it does nothing to ameliorate the
decrease in acceptance rate that accompanies an increase in the large
particle density -- a situation analogous to standard GCE simulations of
single component fluids. Thus even if the effects of the small particles
were to be offset equally for all $N_l$, one would still expect the
transition rate to fall with increasing $N_l$. In such a situation, one
can at best aim to avoid bottlenecks in the sampling by ensuring that (i) is
satisfied {\em locally} in $\{N_l,n\}$. With regard to (ii) and (iii),
there is in practice a tradeoff to be realized here which (in parallel
with satisfying (i)) may necessitate a degree of trial and error,
although more systematic approaches have been considered in the expanded
ensemble literature \cite{Escobedo2007}.  In sec.~\ref{sec:results} we
consider factors affecting the choice for one practical situation.

As discussed in Sec.~\ref{sec:statmech}, the form of $p(N_l,n)$  may
span many decades of probability and in order to sample it effectively,
multicanonical preweighting is called for. This in turn requires
knowledge of a set of weights, $w(N_l,n)$, that facilitate the
even-handed sampling of regions of high and low probability.  One choice
that ensures this is $w(N_l,n) \approx -\log p(N_l,n)$ which results in
a sampled distribution $\hat p(N_l,n)$ that is approximately flat (cf.
Eq.~\ref{eq:weighted})  \cite{footnote3}. However, since $p(N_l,n)$ is
just the distribution that we seek, the task of determining the weight
function appears --at first sight-- to be circular. Fortunately though,
the situation is saved by the observation that it is possible to build up a
suitable estimate of $w(N_l,n)$ from scratch via iterative means
\cite{Berg96}. The approach we favour for doing so is based on the
transition matrix Monte Carlo (TMMC) method
\cite{smith1995,smith1996,Fitzgerald1999,Errington2004,Bruce2003}.

TMMC works by monitoring the transitions between
macrostates and using these to infer their relative probability. Once
sufficient transition statistics have been collected, it is possible to
construct the entire probability distribution. The starting point is the
macrostate balance condition relating the equilibrium probability of two
macrostates $\old$ and $\new$ to the transition rates between them:

\be \label{eq:db}
p(\old) W(\old \rightarrow \new) = p(\new) W(\new \rightarrow \old)\;,
\ee
where $\old$ and $\new$ are taken to represent combinations of $N_l$ and $n$. 
The equilibrium transition rate, $W(\old \rightarrow \new)$ can be estimated in the course
of a simulation by accumulating the acceptance probabilities for
macrostate transitions into a collection matrix, $C(\old \rightarrow \new)$.
For every proposed move, $\old \rightarrow \new$, the {\em unbiased} acceptance probability,
$a$ (calculated from Eq.~\ref{eq:pacc} by assuming $\Delta w=0$) is added to the collection
matrix thus:

\begin{eqnarray}
C(\old \rightarrow \new) &\rightarrow & C(\old \rightarrow \new) + a \\
C(\old \rightarrow \old) &\rightarrow & C(\old \rightarrow \old) + (1-a)\;.
\end{eqnarray}
This happens regardless of whether or not the move is accepted. 

The transition rates can be extracted from the collection matrix via

\be
W(\old \rightarrow \new) = \frac{C(\old \rightarrow \new)}{\sum_{v^\prime} C(\old \rightarrow \new^\prime)}\:,
\ee
where the sum in the denominator on the right hand side runs over all
possible values of the macrovariable.

Putting the transition rates into equation (\ref{eq:db}) yields the 
macrostate probabilities

\be 
\frac{p(\new)}{p(\old)}=\frac{W(\old \rightarrow \new)}{W(\new \rightarrow \old)}\:,
\label{eq:tmprob} 
\ee  
from which the multicanonical weights follow as 

\be 
\label{eq:w8r8}
w(\old) - w(\new) = - \ln \frac{W(\old \rightarrow \new)}{W(\new
\rightarrow \old)}. 
\ee 

Since the collection matrix is concerned solely with unbiased
acceptance probabilities, one is free to apply an arbitrary bias  during
the simulation without affecting estimates of equilibrium
properties. This feature of TMMC can be exploited to provide an
automated strategy for obtaining a suitable multicanonical weight
function. Starting with no knowledge of the weight function, one simply
updates $w(N_l,n)$ periodically via equation (\ref{eq:w8r8}). This allows
the sampling to gradually extend over the range of $N_l,n$, pushing
progressively into regions of ever smaller probability \cite{footnote4}. Once the region of
interest has been adequately sampled, the collection matrix provides an
estimate of the requisite distribution $p(N_l,n)$ via
Eq.~\ref{eq:tmprob}. During the simulation we also sample (in list
form \cite{Wilding2001}) the instantaneous values of $N_l, N_s, n$, together
with the configurational energy $\Phi$. This permits extrapolation of
the results for $p(N_l,n)$ in temperature via standard histogram
reweighting techniques \cite{ferrenberg1989}.

\section{Application to the liquid-vapour transition of a binary Lennard-Jones
mixture}
\label{sec:results}

As a test of our method, we have applied it to the study of liquid-vapour
phase coexistence in a LJ mixture having particle size ratio
$q\equiv\sigma_{ss} / \sigma_{ll}=0.1$ and reservoir volume fraction of
the small particles $\eta^r_s=0.01$. The simulations were performed for
a cubic periodic simulation box of side $L=7.5$, which for this
$\eta^r_s$ would correspond to $N_s\approx 8000$ in the absence of large
particles. Since the coexistence properties of this system are known
already on the basis of simulation studies using a very different
approach (previously proposed by one of us \cite{Liu2006}), there exists
a convenient baseline for comparison. 

The choice of the stage parameters  $\tilde\phi^{(n)}_{\mathrm{min}}$
and $\tilde\phi^{(n)}_{\mathrm{max}}$ was guided by the criteria set out
in Sec.~\ref{sec:implement}. For small values of $N_l\le 130$, only two
intermediate stages were required (ie. $M=3$) to obtain a fairly high
transition rate. However, in order to maintain a roughly constant
transition rate across intermediate stages for different $N_l$, it was
found necessary to vary $\tilde\phi^{(n)}_{\mathrm{max}}$ linearly as a
function of $N_l$ between the limits shown in Table~\ref{tab:levels}.
For $N_l>130$ the overlap of the ghost with large particles becomes the
principal ground for rejecting an insertion, and we chose to mitigate
this by the introduction of an additional stage (assigned to $n=1$) with
parameters
$\tilde\phi^{(1)}_{\mathrm{min}}=\tilde\phi^{(1)}_{\mathrm{max}}=0$,
thus making $M=4$. No variation of the other stage parameters was deemed
necessary in this regime, whose values for
$\tilde\phi^{(n)}_{\mathrm{min}}$ and $\tilde\phi^{(n)}_{\mathrm{max}}$
are included in Table~\ref{tab:levels}. Across the entire range of $N_l$
studied, the acceptance rate for transitions varied from $\simeq 30\%$
at small densities of large particle to $\simeq 5\%$ at liquid-like
densities. The principal source of this variation is overlaps between
the ghost particle and large particles; its magnitude compares favourably
with that occurring in grand canonical studies of single component fluids
over the same density range.

\begin{table}
\begin{tabular}{c|cc|cc}
& \multicolumn{2}{|c|}{$0\leq N_l\leq 130 $} &
\multicolumn{2}{c}{$N_l>130$}  \\
\hline
Stage, $n$ &  $ \tilde\phi_{\mathrm{min}}$&  $
\tilde\phi_{\mathrm{max}}$ &  $ \tilde\phi_{\mathrm{min}}$ &  $
\tilde\phi_{\mathrm{max}}$ \\
1 &  $-0.5$ &  $7.5 \rightarrow 2.7$ & $0$ & $0$ \\
2 &  $-0.8$ &  $20 \rightarrow 16$ & $-0.5$ & $7.5$ \\
3 & -     & - & $-0.8$ & $20$ \\
\end{tabular}
\caption{The stage parameters $\tilde\phi^{(n)}_{\mathrm{min}}$ and
$\tilde\phi^{(n)}_{\mathrm{max}}$ (expressed in units of
$\epsilon_{ls}$) as used in the simulations. For $0\le N_l\le 130$,
two intermediates stages ($n=1,2$) were used (ie. $M=3$), and
$\tilde\phi^{(n)}_{\mathrm{max}}$ was varied linearly as a
function of $N_l$ between the limits shown (see text). For $N_l>130$, 
three intermediate stages were used ($M=4$) with no variation of parameters.}
\label{tab:levels}
\end{table}

The simulations were initialised at the temperature $T^\star=1.047$,
close to the known critical temperature of the model
\cite{Liu2006}. At this temperature the TMMC method was used to obtain a
suitable form for the multicanonical weight function and thence an
estimate of the histogram $p(N_l,n)$ for $N_l=[0:300]$.  This histogram
was then reweighted in $\mu_l$ such as to satisfy the equal area
criterion \cite{Borgs1992} for the two peaks in the near-coexistence
form of $p(N_l)$, thereby yielding an estimate of the coexistence value of
$\mu_l$. Subsequently the data was extrapolated to the lower temperature
$T^\star=1.0$ by means of histogram reweighting. The resulting form of
$p(N_l,n)$ provided a suitable multicanonical weight function for a new
run at this lower temperature, which was again performed for
$\eta^r_s=0.01$ (which necessitated a re-tuning of $\mu_s$). By iterating
this process we were able to step along the coexistence curve without
the need to ever recalculate a multicanonical weight function from
scratch. Further details of this strategy for mapping liquid-vapour
coexistence lines are described in ref.~\cite{Wilding2001}.

\begin{figure}[h]
\includegraphics*[width=\figurewidth]{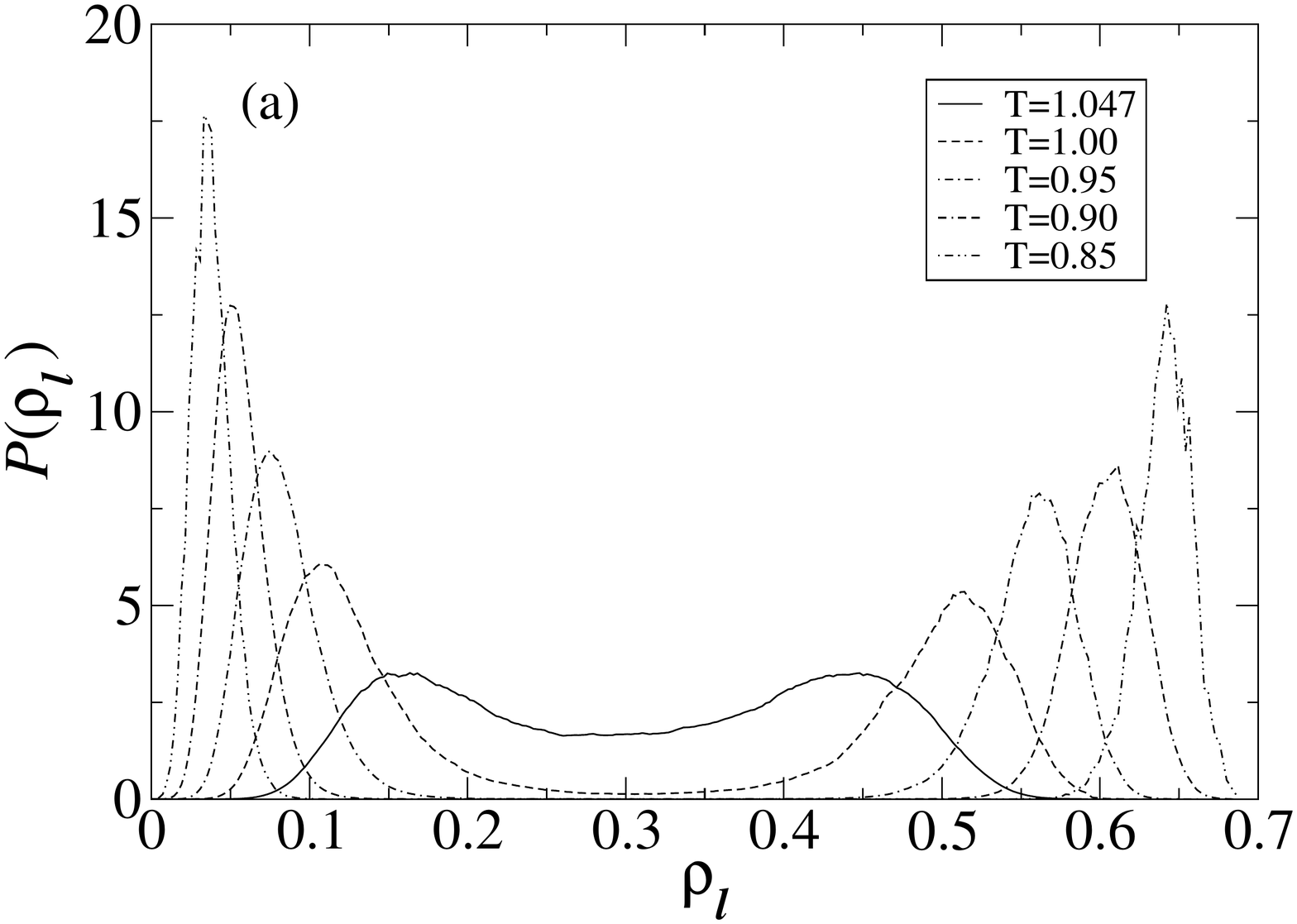}
\includegraphics*[width=\figurewidth]{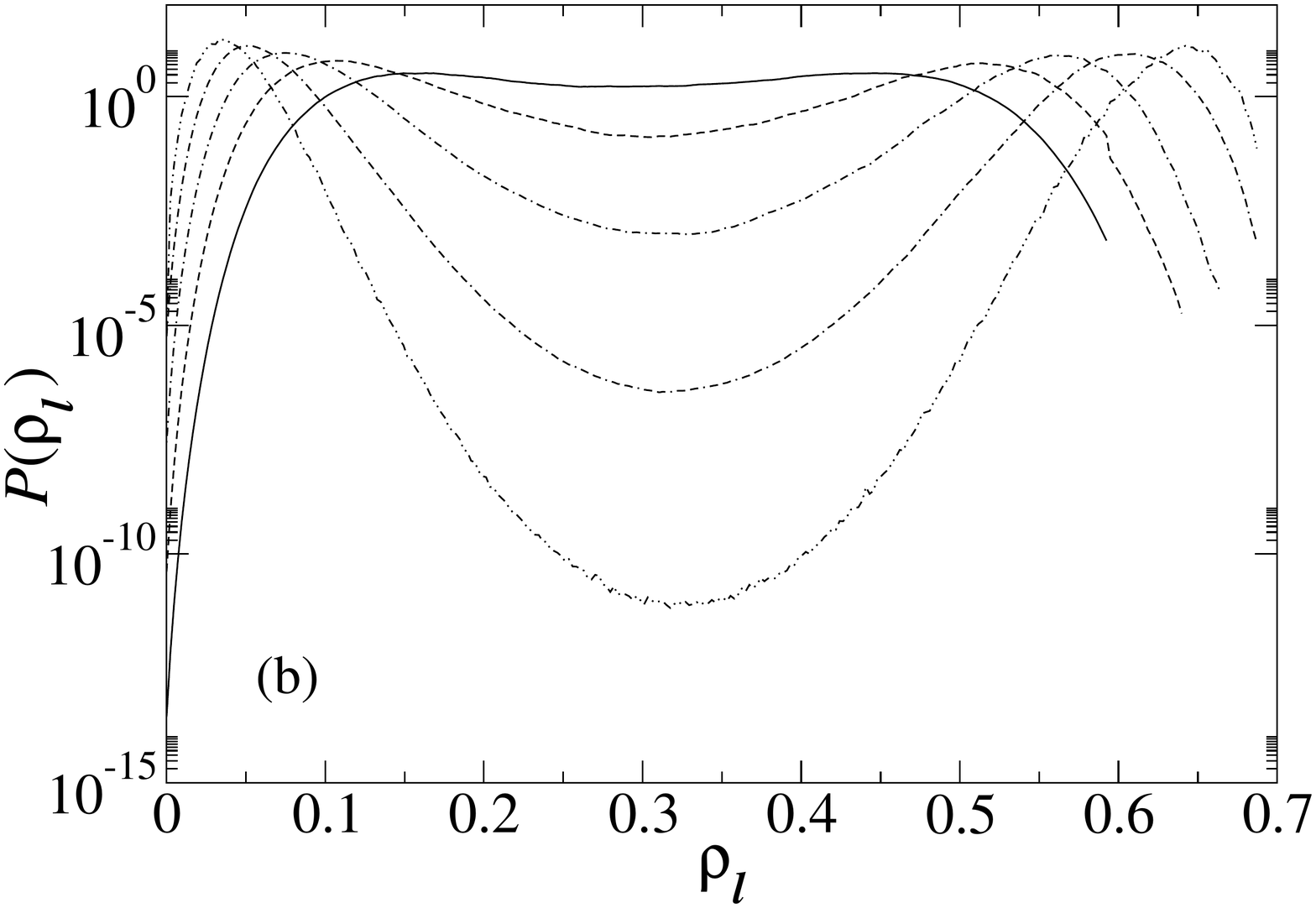}
\caption{{\bf (a)} Estimates of the coexistence forms of $P(\rho_l)$ for
$\eta^r_s=0.01$ obtained using the methods described in the text. Data
are shown for $T^\star=1.047$ (criticality), $1.0, 0.95,
0.90,0.85$. {\bf (b)} The same data expressed on a log scale.}
\label{fig:coexdists}
\end{figure}

\begin{figure}[h]

\includegraphics*[width=\figurewidth]{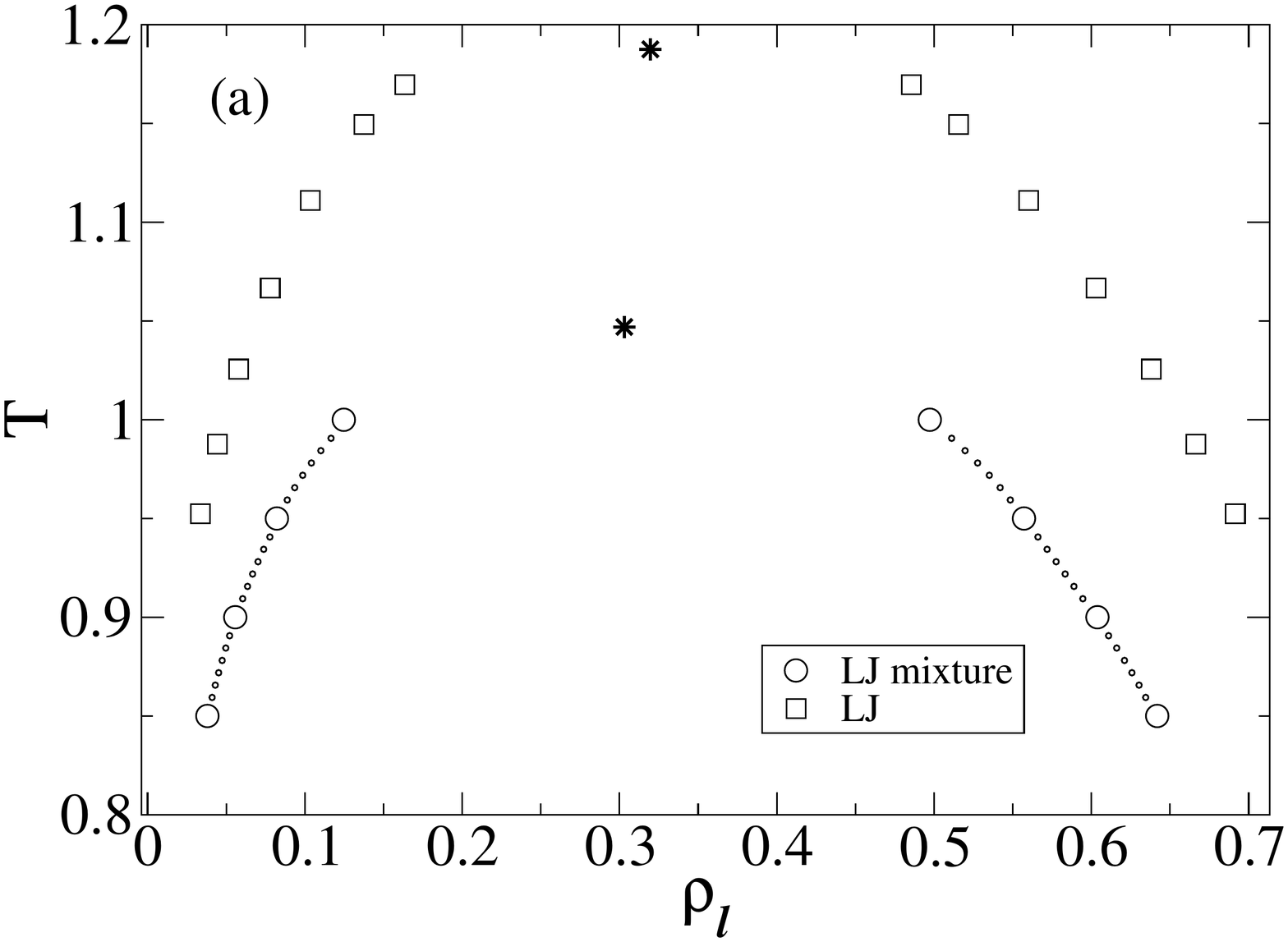}
\includegraphics*[width=\figurewidth]{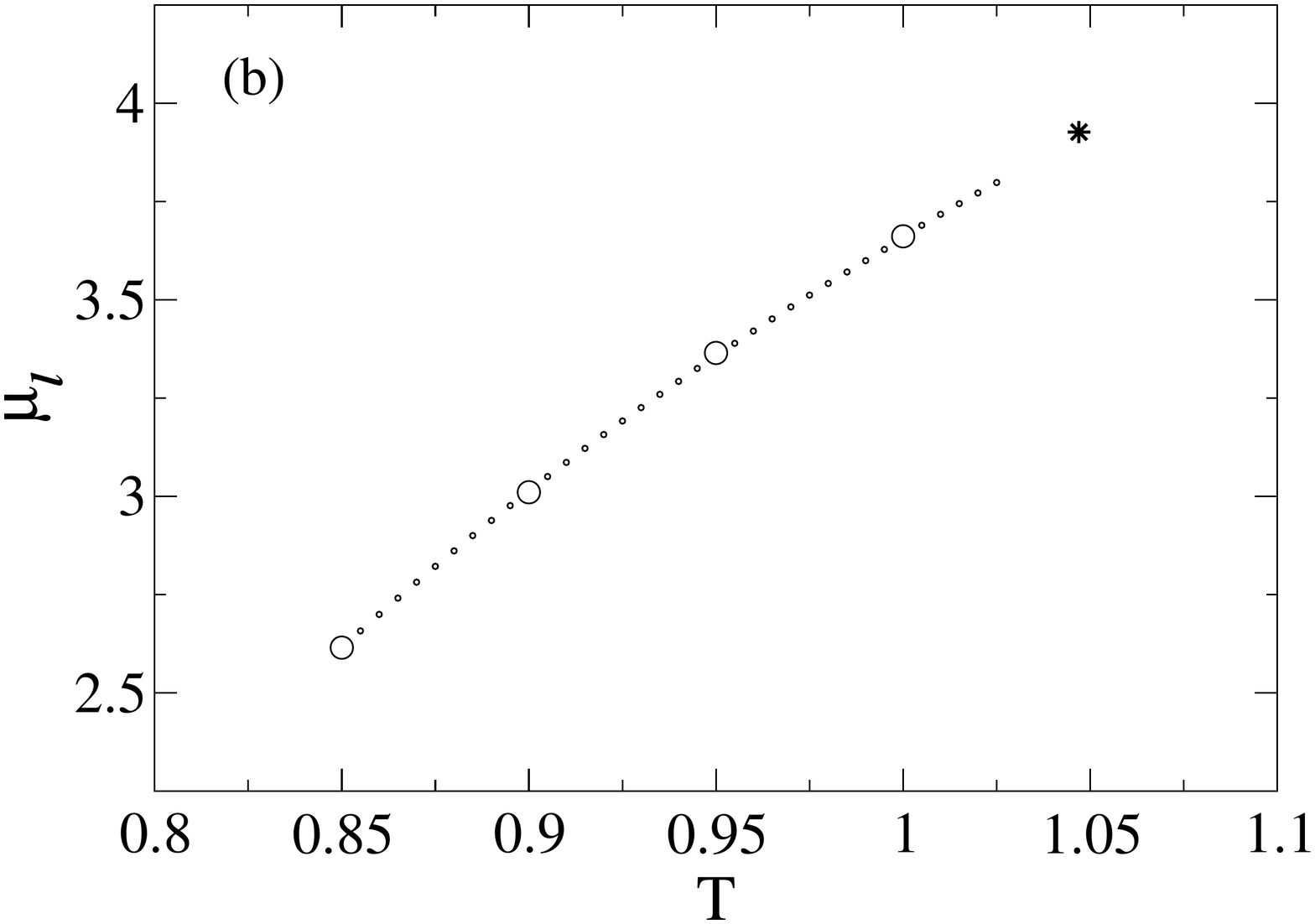}

\caption{{\bf (a}) Coexistence densities (circles) as determined from
the peak positions of Fig.~\protect\ref{fig:coexdists}; dots interpolate
between the measured coexistence densities, and are determined via
histogram reweighting. Squares show the binodal for the single component
LJ fluid obtained in Ref.~\cite{Wilding1995}. Critical points are marked
(*). {\bf (b)} Corresponding coexistence points in the $\mu_l-T$ plane,
with additional points (dots) obtained via histogram extrapolation.}

\label{fig:coexcv}
\end{figure}

Fig.~\ref{fig:coexdists} presents the resulting estimates of the
coexistence forms of $p(\rho_l)$ with ($\rho_l=N_l/V$) at various
temperatures. Not surprisingly, the distributions exhibit behaviour
which is qualitatively similar to that of a single component fluid
\cite{Wilding1995}. An estimate of the corresponding liquid-vapour
binodal can be extracted from the distributions (by averaging the density
under each peak) and is shown in Fig.~\ref{fig:coexcv}(a). The results
are fully consistent with unpublished data (to be presented elsewhere)
obtained using the quite different simulation method of
ref.~\cite{Liu2006}. Also included in Fig.~\ref{fig:coexcv}(a) is the
binodal for the single component LJ fluid determined in a previous study
\cite{Wilding1995}; the comparison reveals that the presence of the
small particles in the mixture depresses the critical temperature
significantly. Estimates of the phase boundary in $\mu_l-T$ space are
shown in Fig.~\ref{fig:coexcv}(b). 

\begin{figure}[h]
  \includegraphics*[width=0.8\columnwidth]{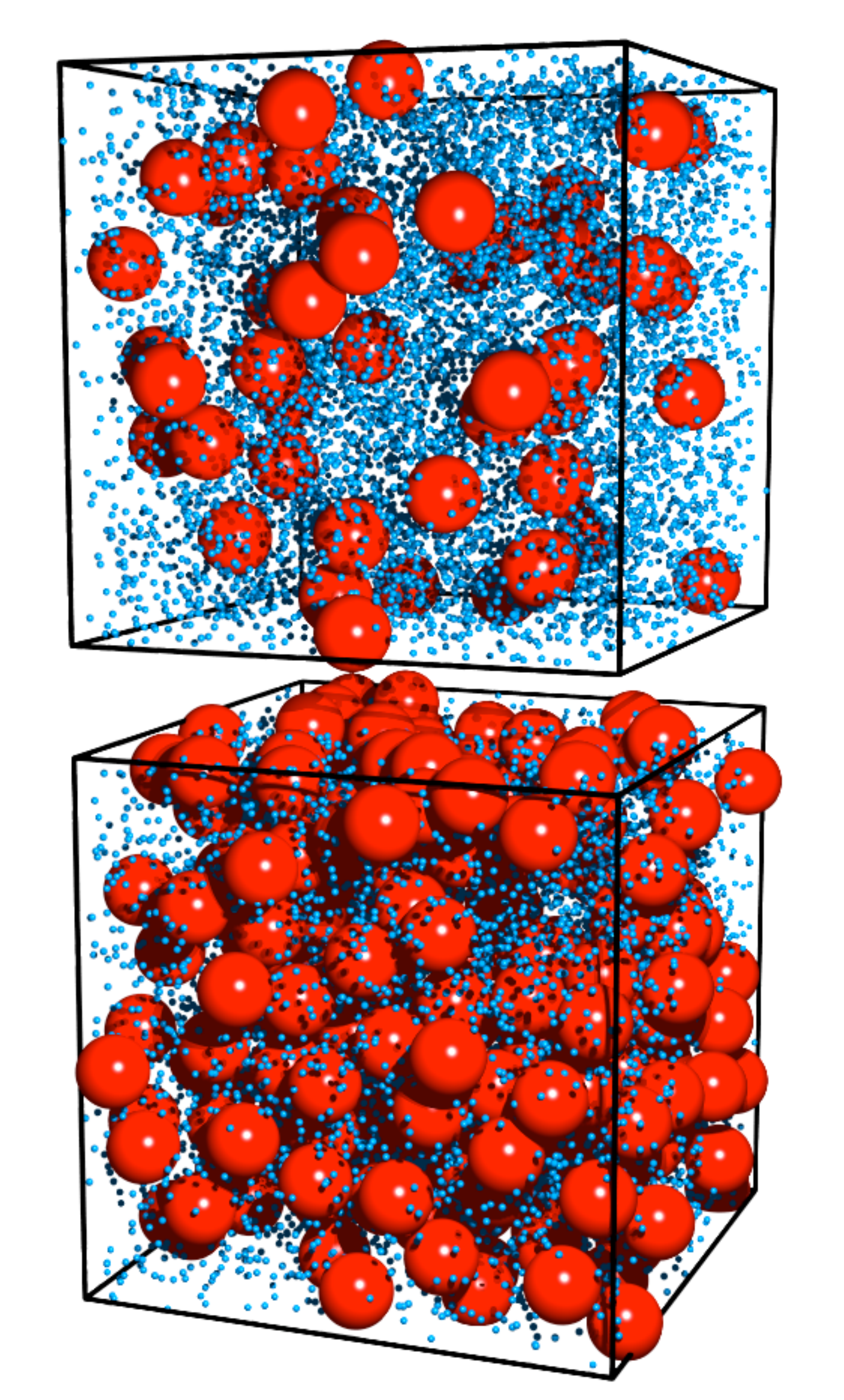} 
\caption{Configuration snapshots of the coexisting vapour 
phase (upper panel) and liquid phase (lower panel) at $T^\star=0.95$.}
\label{fig:snapshot}
\end{figure}

We point out that obtaining this phase diagram in a reasonable
timescale would not have been feasible without the staged insertion/deletion
approach. Our tests show that the wall clock correlation time in the
absence of staging is too large to be reliably estimated. Nevertheless,
a lower bound on the ratio of correlation times with and without staging
can be estimated via a comparison of the transition acceptance rates.
For $\eta_s^r=0.01$, the insertion/deletion rate without staging is
$\sim 10^{-6}$ at liquid-like densities of the large particles. This
very low acceptance rate is of course attributable to the high
likelihood that a randomly chosen large particle insertion results in
overlaps with one or more small particles -- a visual impression of the
difficulty is provided by configurational snapshots of the coexisting
phases as shown in Fig.~\ref{fig:snapshot}. Use of staging increases the
transition acceptance rate to $\sim 10^{-2}$ for $M=4$ stages. The cost
overhead is an increases in the (round trip) random walk length in $N_l$
by a factor of $M$, thereby increasing the correlation time by a factor
$M^2~\sim 10$. Hence we believe that in the present case our method is
more efficient than standard grand canonical sampling by a net factor of
$\sim 10^3$.  

Notwithstanding the impressive scale of this speedup, the net
computational expenditure incurred by our study remained significant.
This is primarily due to the large number of small particles in the
system, even for the relatively low volume fractions of small particles
that we considered. To be more quantitative, the task of obtaining the
initial multicanonical weight function consumed about a week of CPU time
on a 32-core 3 GHz machine, while data collection for each subsequent
coexistence state point also took about a week.

\section{Conclusions}

In summary, we have described a grand canonical Monte Carlo simulation
scheme for the study of fluid phase transitions in highly
size-asymmetrical binary mixtures. The method overcomes the low
acceptance rate for large particle transfers that plagues standard GC
approaches. This is achieved via a staged insertion scheme in which
insertion (deletion) of a large particle proceeds stochastically via a
set of intermediate states in which the coupling to the environment of
small particles is switched on (off) gradually in stages. Once a
suitable set of stages and associated multicanonical weights has been
determined, the system essentially performs a random walk in the density
of the large particles. We have applied the method to a particular
binary Lennard-Jones mixture having $q=0.1$ and $\eta_s^r=1\%$,
determining the coexistence envelope for liquid-vapour demixing of the
large particles.

As regards the outlook for this approach, we see no reason why it
shouldn't be effective at larger reservoir volume fractions of the small
particles, or indeed for multicomponent mixtures. The principal
computational overhead associated with higher values of $\eta^r_s$ will
be the larger number of interactions with small particles. The number of
stages $M$ necessary to maintain a reasonable acceptance rate will
presumably increase too.  We intend to investigate and report on these
issues in future work.


\acknowledgments 

It is a pleasure to contribute to this Special Issue of {\em
Molecular Physics} celebrating the work of Professor Bob Evans. During
his career, Bob has made numerous seminal contributions to liquid state
theory, been a tireless champion of the field, and an inspiration to those
in it. We wish him many rewarding years to come. This work
was supported by EPSRC grant EP/F047800. Computational results were
partly produced on a machine funded by HEFCE’s Strategic Research
Infrastructure fund.

\bibliography{/Users/pysnbw/Papers}

\begin{thebibliography}{10}
\bibitem{footnote1} Note that this volume fraction is notional in the
sense that we use the value of $\sigma$ as if it were a hard core radius
$\eta_s=\pi N_s\sigma_s^3/(6V)$ where $V$ is the system volume.
\bibitem{footnote2}We note in passing that further efficiency
gains accrue by reducing the cell side by a factor of 2 and summing over
a greater number of cells, thereby reducing the volume to be searched by
up to a factor of $2$.
\bibitem{footnote4} In practice, the initial estimate of $w(N_l,n)$ can
be obtained more rapidly by restricting the range of $N_l$ that can be
sampled in a given run. This is done by holding $N_l$ inside a fixed
window, $N_l^{\mathrm{low}} \le N_l \le N_l^{\mathrm{high}}$. The
results from different windows can be combined self consistently by
simply merging the collection matrix from each.
\end{thebibliography}
\bibliographystyle{prsty}

\end{document}